\documentclass[doublespacing]{elsart}

\usepackage{graphicx}
\usepackage{amssymb}
\usepackage{bm}

\begin{document}

\begin{frontmatter}

\title{Resonant helical deformations in nonhomogeneous Kirchhoff filaments}

\author[if1]{Alexandre F. da Fonseca,}
\author[if1]{C. P. Malta\corauthref{a1}}
\ead{coraci@if.usp.br}
\corauth[a1]{C. P. Malta}
\and
\author[if2]{M. A. M. de Aguiar}

\address[if1]{Instituto de F\'{\i}sica, Universidade de S\~ao Paulo,
USP\\ Caixa Postal 66318, 05315-970, S\~ao Paulo, Brazil}

\address[if2]{Instituto de F\'{\i}sica `Gleb Wataghin', Universidade 
Estadual de Campinas, \\ UNICAMP 13083-970, Campinas, SP, Brazil}  


\begin{abstract} 

We study the three-dimensional static configurations of nonhomogeneous
Kirchhoff filaments with periodically varying Young's modulus. This
type of variation may occur in long tandemly repeated sequences of
DNA. We analyse the effects of the Young's modulus frequence and
amplitude of oscillation in the stroboscopic maps, and in the regular
(non chaotic) spatial configurations of the filaments.  Our analysis
shows that the tridimensional conformations of long filaments may
depend critically on the Young's modulus frequence in case of
resonance with other natural frequencies of the filament. As expected,
far from resonance the shape of the solutions remain very close to
that of the homogeneous case. In the case of biomolecules, it is well
known that various other elements, besides sequence-dependent effects,
combine to determine their conformation, like self-contact, salt
concentration, thermal fluctuations, anisotropy and interaction with
proteins. Our results show that sequence-dependent effects alone may
have a significant influence on the shape of these molecules,
including DNA. This could, therefore, be a possible mechanical
function of the ``junk'' sequences.
\end{abstract}

\end{frontmatter}


The study of tridimensional structures of filamentary objects is of
great interest in several areas of knowledge, ranging from microscopic
to macroscopic systems. Examples of macroscopic systems in Engineering
are the stability of suboceanic cables~\cite{coyne,zajac} and
installation of optical fibers~\cite{sun,vaz}; in Biology, the shape
of climbing plants~\cite{alain}; and, in Physics and Mathematics, the
shape and dynamics of cracking whips~\cite{alain2}. The microscopic
examples are in the area of the Structural Biology, as in the study of
the structure of biomolecules~\cite{tamar,yang} and bacterial
fibers~\cite{wolge,klaper}.

Filamentary systems are usually modeled as thin {\it uniform} rods.
However, nonuniformities in the filament properties can affect
significantly its tridimensional structure, so in this work we study
certain resonant variations in the shape of helical rods induced by
periodic variations in the stiffness of the rod. One of the
motivations for this theoretical study is, on one hand, the
observation that the DNA stiffness is
sequence-dependent~\cite{olson}. On the other hand, it is also
known~\cite{jenny,brian} that a substantial fraction of all eukaryotic
genomes are composed of tandemly repeated sequences of
base-pairs. These repetitive DNAs are formed by nucleotide sequences
of varying length and composition reaching up to 100 megabasepairs of
length~\cite{brian}. Usually they are regarded as ``selfish'' or
``junk'' DNA~\cite{mclister} because they seem to have little or no
functional role. In fact, some studies suggest that the behavior of
repetitive sequences can be, in some cases, beneficial to the organism
and, in others, harmful \cite{jenny,brian}. It could also be related
to some form of cancer \cite{gabriele}.

In this work we consider rods with small periodic variation of the
Young's modulus, motivated by the existence of a large number of
repetitive sequences of DNA. So the numerical calculations presented
here have been performed using DNA parameters. Nevertheless, the
qualitative results remain valid for general rods with periodic
stiffness variation. We remark that ideal elastic rod models are not
considered to give realistic solutions for the spatial structure of
the DNA or other biomolecules~\cite{wilma}. Therefore, our results for
the tridimensional configurations must be considered as general
mechanical tendencies due to sequence-dependent variations of the
Young's modulus, rather than exact solutions for the DNA
structure. Realistic models for the DNA consider base-pair parameters,
as the recent theory of sequence-dependent DNA elasticity proposed by
Coleman, Olson and Swigon~\cite{col3}, where the elastic energy
depends on a function of the six kinematical variables relating the
relative orientation and displacement between successive
base-pairs. Thermal fluctuations play also an important role in the
structure of these molecules and, therefore, statistical mechanical
approaches are more appropriate to model their spatial
configurations~\cite{mezard,anselmi,rabin}. Nevertheless, in order to
analyse the effect of periodic nonhomogeneities in equilibrium
configurations, it is sufficient to take into account thermal
fluctuations just roughly. This has been done by adopting an excess of
5\% of linking number~\cite{marko2}.

A table with the Young's modulus of all 32 trinucleotide units was
recently obtained in~\cite{gromiha} that can be used as a reference
for the amplitude of the variation. We analyse the effects of the
frequency of these periodic modulations in the Young's modulus, in
terms of dynamical stroboscopic maps and directly in the
tridimensional structure of the rod. We are particularly interested
in understanding how the sequence-dependent mechanical properties of
the filament can cause its shape to deviate from the well known
uniform solution, namely, the helix.

Mielke and Holmes~\cite{holmes} demonstrated that the variation of the
bending stiffness along the rod can cause spatially complex
tridimensional shapes and sensitivity with respect to initial
conditions. They described {\it infinitely} long rods as initial value
problems (IVP) and used analytical techniques based on dynamical
systems theory to study some specific hyperbolic fixed points related
to homoclinic orbits. 

It should be stressed that here we analyze a case where it is not
possible to use the perturbative methods of dynamical systems theory,
therefore we had to resort to numerical simulations.

The Kirchhoff rod model has been extensively used in the literature to
model continuous rods~\cite{kirch,dill,tabor}. For example, Shi and
Hearst~\cite{shi} and Nizette and Goriely~\cite{nize} obtained and
classified, respectively, all the solutions of the static Kirchhoff
equations for homogeneous rods with circular cross section. Coleman
{\it et al}~\cite{col1} made a complete analysis of the stability of
DNA within the framework of the Kirchhoff rod model, subjected to
strong anchoring end conditions. Recently, da Fonseca and de
Aguiar~\cite{fonseca1} compared homogeneous and nonhomogeneous rods,
with varying stiffness, subjected to boundary conditions. The effects
of nonhomogeneous mass distribution in the dynamics of unstable closed
rods have been analyzed by Fonseca and de Aguiar~\cite{fonseca2}. Yang
\emph{et al}~\cite{yang} have studied a particular case of
nonhomogeneous Young's modulus for closed rods. Manning \emph{et
al}~\cite{madoc} have incorporated into the Kirchhoff model the
sequence-dependent discrete data of the intrinsic curvature of DNA. In
the present study we assume that the DNA is intrinsically an untwisted
straight rod, but we include sequence-dependent stiffness. 

In the Kirchhoff model for an inextensible rod the Young's modulus
appears in the equations through its {\it bending} coefficient. As
there is no table with the {\it bending} coefficients for all {\it di}
or {\it trinucleotides}, we shall consider the data in \cite{gromiha}
as a reference for our calculations, as mentioned.

We shall consider the Hamiltonian formulation of the Kirchhoff's
equations.  We shall follow the simplest derivation by Nizette and
Goriely~\cite{nize} instead of the rigorous Hamiltonian formulation by
Mielke and Holmes \cite{holmes}. The reader is referred
to~\cite{dill,tabor,fonseca2} for a derivation of the Kirchhoff model,
and to ref.~\cite{nize} for the Hamiltonian formulation. Hamilton's
equations for the Kirchhoff model are analogous to those of a
symmetric spinning top in a gravity field, with the arc length \( s \)
along the rod playing the role of time. The main advantage of a
Hamiltonian formulation is that the theory of chaotic Hamiltonian
systems and stroboscopic maps can be directly applied to understand
the spatial behavior of the filament. The length of the tandemly
repeated sequences can reach up to 100 megabasepairs \cite{brian}
while the length of the repeats is no more than a few hundreds of
basepairs long. So we shall solve the Kirchhoff's equations as an
initial value problem (IVP) to find the conformational solutions of
the filament problem.

The Hamiltonian for an elastic rod with circular cross section, in
Euler angles, is
\begin{equation}
\label{energia}
H=\frac{P^{2}_{\theta }}{2E}+\frac{P^{2}_{\phi }}{2\Gamma _{0}\mu}+
\frac{\left( P_{\psi }-P_{\phi }\cos \theta \right) ^{2}}{2E\sin ^{2}\theta }
+F\cos \theta ,
\end{equation}
where \( E=E(s) \) and \( \mu =\mu (s) \) are the scaled Young's and
shear moduli, respectively. \( F \) is the intensity of the total
contact force (constant) exerted on the cross section at $s$. The
units used here are the same of refs. \cite{dill,tabor,fonseca2}. \(
\Gamma _{0} \) varies between \( 2/3 \) (incompressible material) and
\( 1 \) (hyper-elastic material), and it should be remarked that
$\Gamma _0$ has no influence in the equilibrium solutions. The momenta
are defined by
\begin{equation}
\label{Pteta}
\begin{array}{ll}
P_{\theta }&\equiv  E(s)\theta ' \; ,\\
P_{\phi }&\equiv  \Gamma _{0}\mu (s)\left( \phi '+\psi '\cos \theta
\right) \; ,\\
P_{\psi }&\equiv E(s)\psi '\sin ^{2}\theta +P_{\phi }\cos \theta \; ,
\end{array}
\end{equation}
where the prime indicates the derivative with respect to $s$. 

In the case of a homogeneous filament, $E(s)\equiv1$ and $\mu
(s)\equiv1$ in the equation (\ref{energia}), and the Hamiltonian is
written as:
\begin{equation}
\label{homoene}
H=\frac{P^{2}_{\theta }}{2}+\frac{P^{2}_{\phi }}{2\Gamma _{0}}+
V(\theta) ,
\end{equation}
where the potential $V(\theta)$ is:
\begin{equation}
\label{vteta}
V(\theta)=\frac{\left( P_{\psi }-P_{\phi }\cos \theta \right)
^{2}}{2\sin ^{2}\theta } +F\cos \theta .
\end{equation}
If $P_{\psi}= P_{\phi}$, then $\theta=0$ (straight rod) is an equilibrium
solution, and if $F\not=0$  there is a second equilibrium solution
($\theta\not=0$) corresponding to a helix~\cite{vander}. If
$P_{\psi}=- P_{\phi}$, the equilibrium solution is $\theta=\pi$, and
if $F\not=0$ there is a second equilibrium solution as above. If
$|P_{\psi}| \neq |P_{\phi}|$ then $V(\theta)$~(\ref{vteta}) has a
single minimum corresponding to the well known helix solution.
Denoting by $\theta_{0}$ the point of minimum of the potential
(\ref{vteta}), the frequency $\omega_0$ of small oscillations around
$\theta_0$ is given by:
\begin{equation}
\label{freq0}
\omega_{0}=P_{\phi}^2+2V(\theta_{0})-6F\cos \theta_{0}. 
\end{equation}
In the Appendix we show that the potential $V(\theta)$~(\ref{vteta})
cannot be well approximated by a second order expansion in the neigborhood
of its minimum (it is expanded up to order 6).

The Euler angles $\theta$, $\phi$ and $\psi$ connect a fixed Cartesian
basis $\{{\bf{e}}_{1},{\bf{e}}_{2},{\bf{e}}_{3}\}$ to the \emph{local
orthonormal basis} ${\bf{d}}_{i}={\bf{d}}_{i}(s,t)$, $i=1,2,3$,
attached to each point of the rod. The direction of ${\bf{e}}_3$ is
chosen to be the direction of the constant force $\mathbf{F}$.
${\bf{d}}_{3}$ is chosen to be tangent to the curve ${\bf{x}}(s)$ that
defines the axis of the filament, and ${\bf{d}}_{1}$ and
${\bf{d}}_{2}$ are in the direction of the principal moments of
inertia of the cross section (perpendicular to ${\bf{d}}_{3}$). The
momentum $P_{\psi}$ is the ${\bf{e}}_{3}$-component of the angular
momentum with respect to the axis of the rod and $P_{\phi}$ is the
torsional moment, {\it i. e.}, the momentum with respect to
${\bf{d}}_{3}$ \cite{nize}. They remain constant along the rod even if
\( E \) and \( \mu \) depend on the arc length \( s \). The
Hamiltonian, eq.(\ref{energia}), will depend on $s$ through $E(s)$ and
$\mu (s)$.

We shall consider the following periodic variation of the scaled
Young's modulus: 
\begin{equation}
\label{young}
E(s)=1+\alpha \cos \omega s \;.
\end{equation}
The parameter $\alpha$ is the amplitude of the Young's modulus
periodic variation, and $\omega$ is the frequency of the
oscillation. We are concerned only with the the tridimensional shape
of the rod and it should be stressed that the shear modulus \( \mu (s)
\) does not affect the tridimensional configuration.

To obtain the equilibrium configurations we first solve Hamilton's
equations for \( \theta \) and \( P_{\theta } \). Then, we solve
Eq.(\ref{Pteta}) for \( \psi \) and reconstruct the filament by
integrating ${\bf{d}}_{3}$ along $s$:
\begin{equation}
\label{xs}
{\bf{x}}(s) =\int _{0}^{s}\; [(\sin {\theta }\cos {\psi
})\ {\bf{e}}_{1}+(\sin {\theta }\sin {\psi })\ {\bf{e}}_{2}+(\cos {\theta
})\ {\bf{e}}_{3}]ds^{\prime }
\end{equation}  

${\bf{x}}(s)$ is a function of the initial conditions \( \theta
(s=0)\equiv \theta _{0} \) and \( P_{\theta }(s=0)\equiv P_{0}
\). Without lack of generality, \( P_{0} \) can be set equal to \( 0
\) so that \( \theta _{0} \) will be a conformation parameter. In
solving the equation for \( \psi \) we set its initial value \( \psi
_{0}=0 \).

In what follows we present numerical calculations performed with the
following fixed mechanical parameters: $\alpha=0.1$, $P_{\psi}=0.086$,
$P_{\phi}=0.043$ and $F=20$pN. These parameters, excepting the force
$F$, are written in properly scaled units. The maximum value for
$\alpha$ is $0.66$, in accordance with the table of the DNA Young's
modulus presented in reference~\cite{gromiha}. The value of $P_{\phi}$
used corresponds to an excess of \( 5\% \) of the linking
number~\cite{marko2} due to thermal fluctuations. The value of the
force corresponds to a compressing force consistent with the values in
the literature~\cite{macgregor}.

Fig~\ref{fig1} displays nine stroboscopic maps on the
$\theta-P_{\theta}$ plane for different values of frequency
$\omega$. We start with $\omega=0.60\omega_{0}$ (Fig~\ref{fig1}a)
where a larger stability island encloses the main equilibrium point at
$\theta\simeq2.08$rad and $P_{\theta}=0$, and a smaller island is seen
on the left, at $\theta\simeq 0.5$rad. The frequency goes up to
$\omega=2.00\omega_{0}$. We recall that spatial chaos has been
observed before in the Kirchhoff equations~\cite{holmes,davies}.

As we go through the sequence of stroboscopic maps displayed in the
Fig~\ref{fig1}, the large island in the plate (a) slowly shrinks
and eventually disappears at $\omega \approx \omega_0$
(Figs~\ref{fig1}e and~\ref{fig1}f). A second important island appears
at $\omega \approx 0.82\omega_0$ (Fig~\ref{fig1}b). This new island
increases in size and moves towards the right as $\omega$ is
increased. Besides these two main islands, a number of smaller and
short-lived islands pop up and disappear as $\omega$ changes, a
phenomenon typical of chaotic maps. We shall concentrate our study on
the two main islands described above, since they dominate the
stroboscopic maps and last for large intervals of $\omega$.

We shall now investigate the differences in the shape of the
tridimensional configurations corresponding to the two equilibrium
points lying at the center of these islands. In order to construct the
rods we solved the Hamiltonian equations using the values of the
equilibrium point for $\theta$ and $P_{\theta }$ as initial conditions
$\theta_{0}$ and $P_{0}$ and used the equation (\ref{xs}) to construct
the filament.

The tridimensional configuration corresponding to the equilibrium
point changes as the frequency is varied. The shape evolution is
displayed in Figs~\ref{fig2} and~\ref{fig3} for the two main
equilibrium points mentioned above.

In Fig~\ref{fig2}, panels (a) to (d), we show the shape evolution
of the configuration corresponding to the first main equilibrium point
which lies in the center of the main island appearing in the
Fig~\ref{fig1}a, and in the center of the island on the right in
the Figs~\ref{fig1}b,~\ref{fig1}c and~\ref{fig1}e, respectively. We
can see that the shape of the rod deviates more and more from the
helix pattern as $\omega$ is increased, becoming rather twisted for
$\omega=0.92\omega_0$.

In Fig~\ref{fig3}, panels (a) to (d), we show the shape evolution
of the configuration corresponding to the other main equilibrium point
which lies in the center of the `left island' (born in the
Fig~\ref{fig1}b). The configurations shown in the Figs~\ref{fig3}a-d
correspond to the frequency values use in the Figs~\ref{fig1}b,
\ref{fig1}c, \ref{fig1}f and \ref{fig1}h, respectively. The behavior
of this sequence of rod shapes is the reverse of that corresponding to
the first equilibrium point (Fig~\ref{fig2}). As $\omega$ increases,
the shape becomes less coiled and eventually recovers the near-helix
shape, similar to the rod in Fig~\ref{fig2}a, corresponding to the
first equilibrium point (Fig~\ref{fig1}a).

Finally, when $\omega=2 \omega_0$, Fig~\ref{fig1}i, a
`period-doubling' bifurcations occurs. The orbit at the center of the
island becomes unstable and a new stable equilibrium, with twice the
original period, appears. 

The sensitivity of the shape of the nonhomogeneous rods to the
amplitude of the nonhomogeneity can also be tested. Fig~\ref{fig4}
shows, for the same mechanical parameters of the previous figures, the
helix solution of the homogeneous case (left), the solution for
$\alpha=0.001$ (middle) and the solution for $\alpha=0.01$ (right), in
the resonant case $\omega=\omega_0$. Also, these solutions can be
compared to that in the Fig~\ref{fig3}c. We can see that even for very
small values of the amplitude $\alpha$, the tridimensional
configuration deviates fast from the helix solution at the resonance.

It is interesting to notice that, as the frequency increases, the
position of the equilibrium points move in the direction of increasing
$\theta$. Fig~\ref{fig5} displays the value of $\theta$ corresponding
to the equilibrium point related to the main~(circles) and to the
left~(square) islands as function of $\omega$. The dotted line
indicates $\theta_0$ which is the position of the equilibrium point of
the potential $V(\theta)$ related to the homogeneous case. As we can
see in the Figs~\ref{fig2}a and \ref{fig3}d, the shape of the
corresponding tridimensional configuration becomes similar to the
helix when the position of the equilibrium position gets close to
$\theta_0$~(homogeneous case equilibrium point position).
 
The main result of this numerical experiment is that the
tridimensional conformations of long filaments may depend critically
on sequence-dependent properties if these are in resonance with other
natural periods of the filament. As expected, in the limit of very low
or very high frequencies, as compared to $\omega_{0}$, the shape of
the solutions remain very close to that of the homogeneous case. In
the case of biomolecules, it is well known that various other
elements, besides sequence-dependent effects, combine to determine
their conformation, like self-contact, salt concentration,
thermal fluctuations, anisotropy and interaction with proteins. Our
results show that sequence-dependent effects alone may have a
significant influence on the shape of these molecules, including
DNA. This could, therefore, be a possible mechanical function of the
``junk'' sequences. 
 
This work was partially supported by the Brazilian agencies FAPESP,
CNPq and FINEP.

\appendix

\section*{Appendix}

Here we show that the potential $V(\theta)$ of
equation (\ref{vteta}) cannot be well approximated by an expansion up to
order 2 or 3 around its minimum at $(\theta=\theta_0)$. This is due to the
presence of $\sin^{2}\theta$ in the denominator of one of the terms of
the  $V(\theta)$. To illustrate this, we expand $V(\theta)$, 
up to order 6 in $(\theta-\theta_0)$, for the same numerical
parameters used in this paper,

\begin{equation}
\label{taylor}
\begin{array}{l}
V(\theta)=V(\theta_0)+0.0205(\theta-\theta_0)^2+
0.0227(\theta-\theta_0)^3+0.0235(\theta-\theta_0)^4+ \\
0.0257(\theta-\theta_0)^5+
0.0275(\theta-\theta_0)^6+O[(\theta-\theta_0)^7] \; ,
\end{array}
\end{equation}
where $\theta_0\simeq2.043$ for this case.

\begin{figure}[ht] 
  \begin{center}
  \includegraphics[height=45mm,width=45mm,clip]{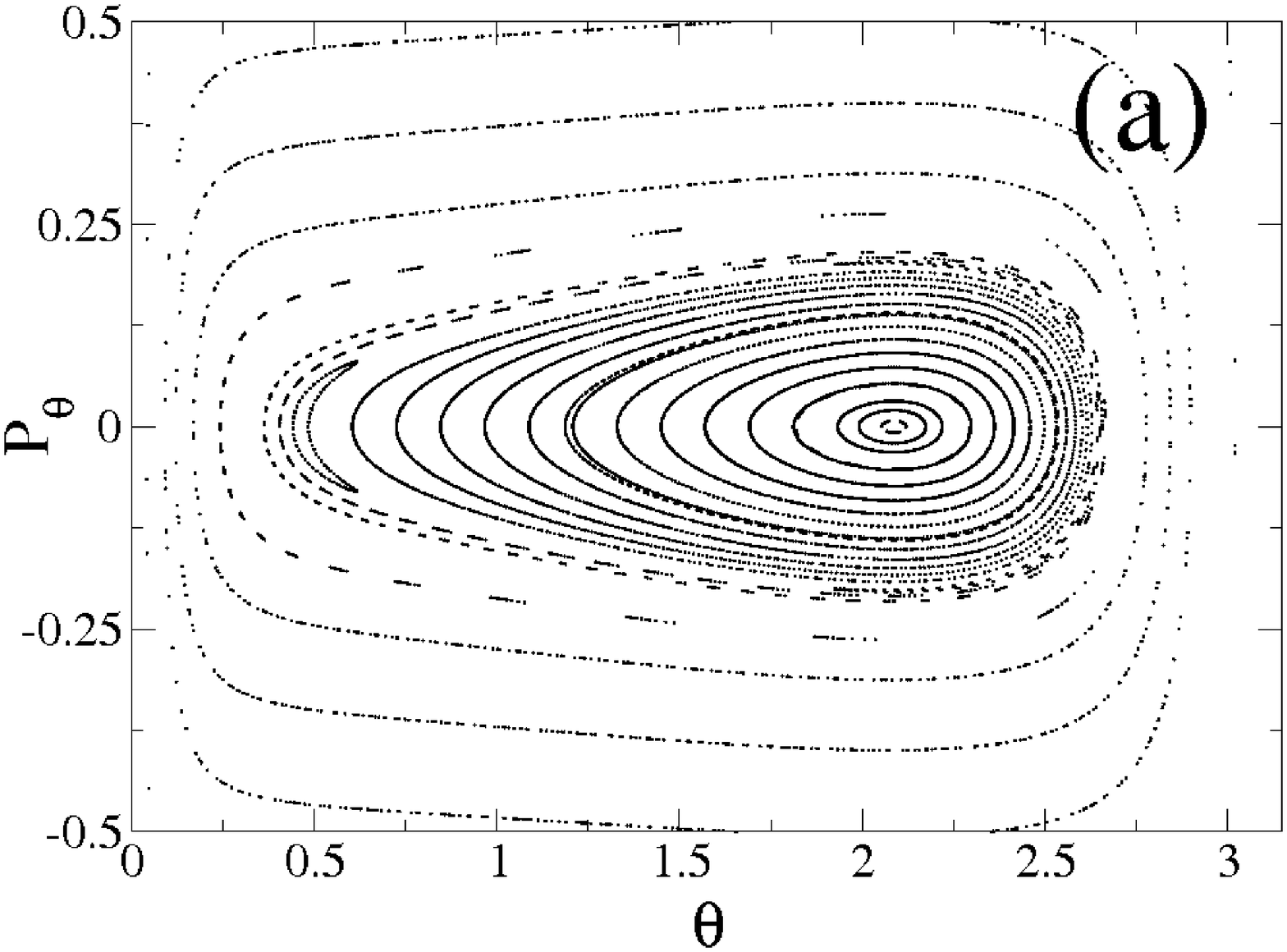}
  \includegraphics[height=45mm,width=45mm,clip]{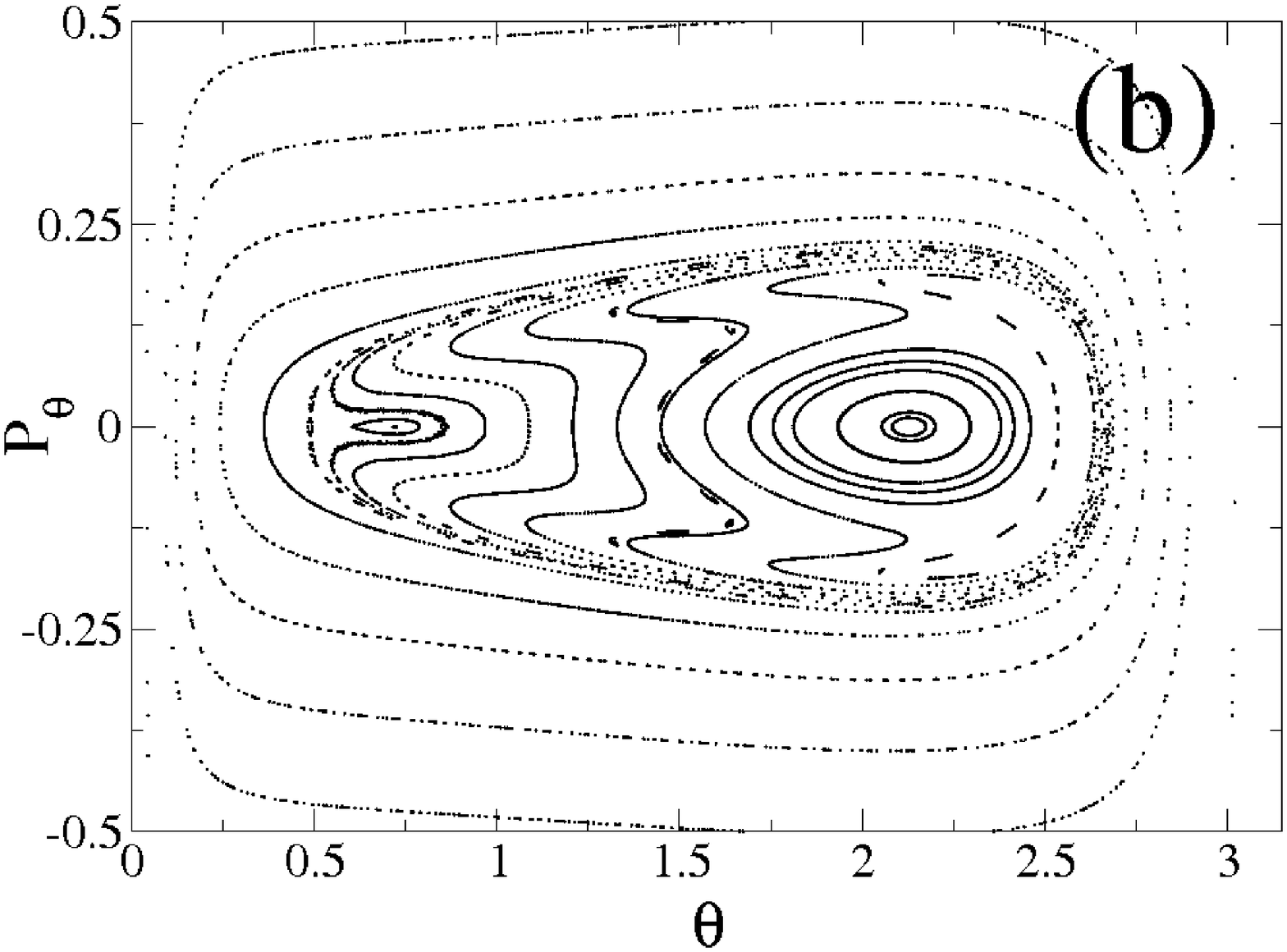}
  \includegraphics[height=45mm,width=45mm,clip]{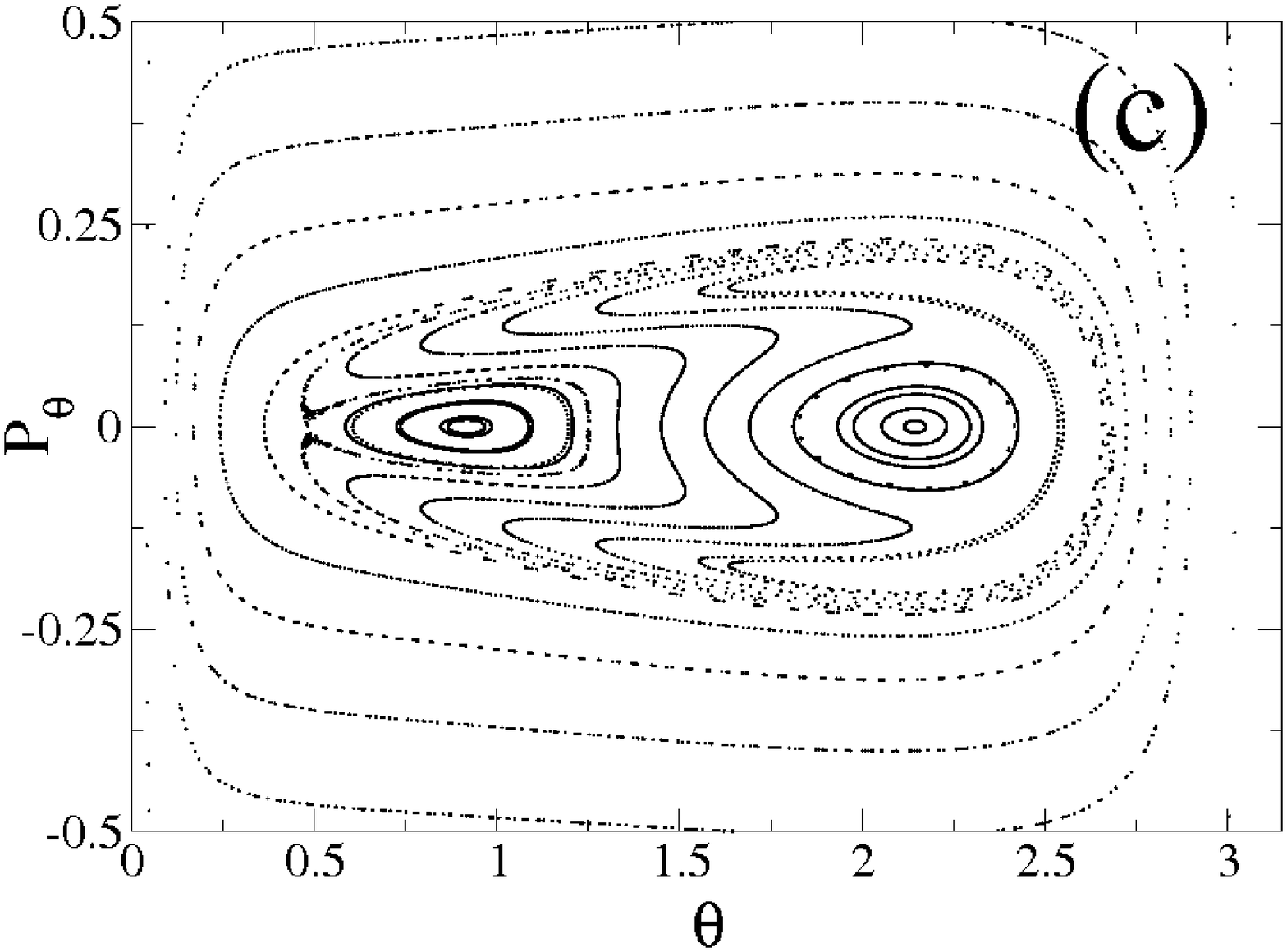}
  \includegraphics[height=45mm,width=45mm,clip]{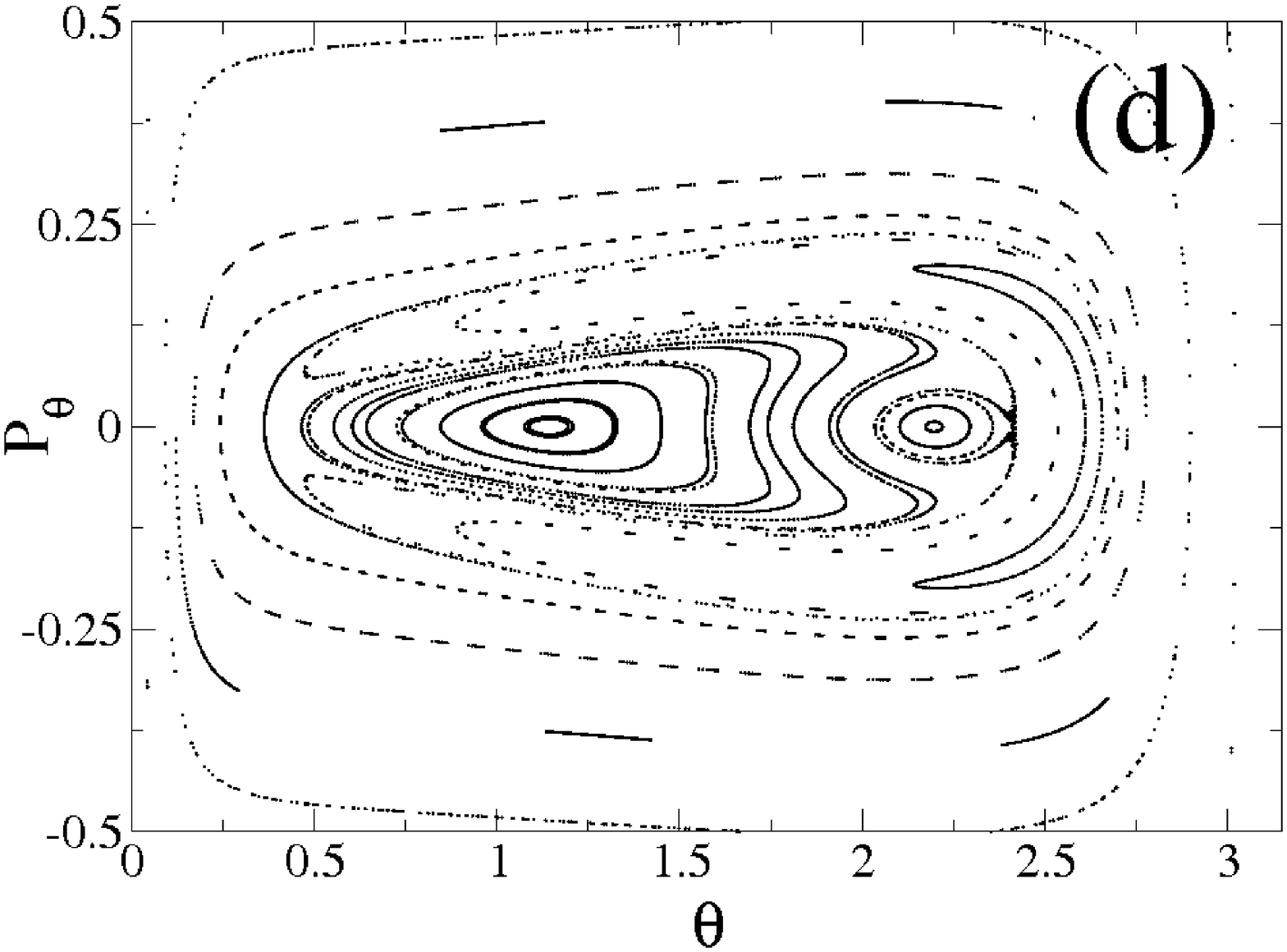}
  \includegraphics[height=45mm,width=45mm,clip]{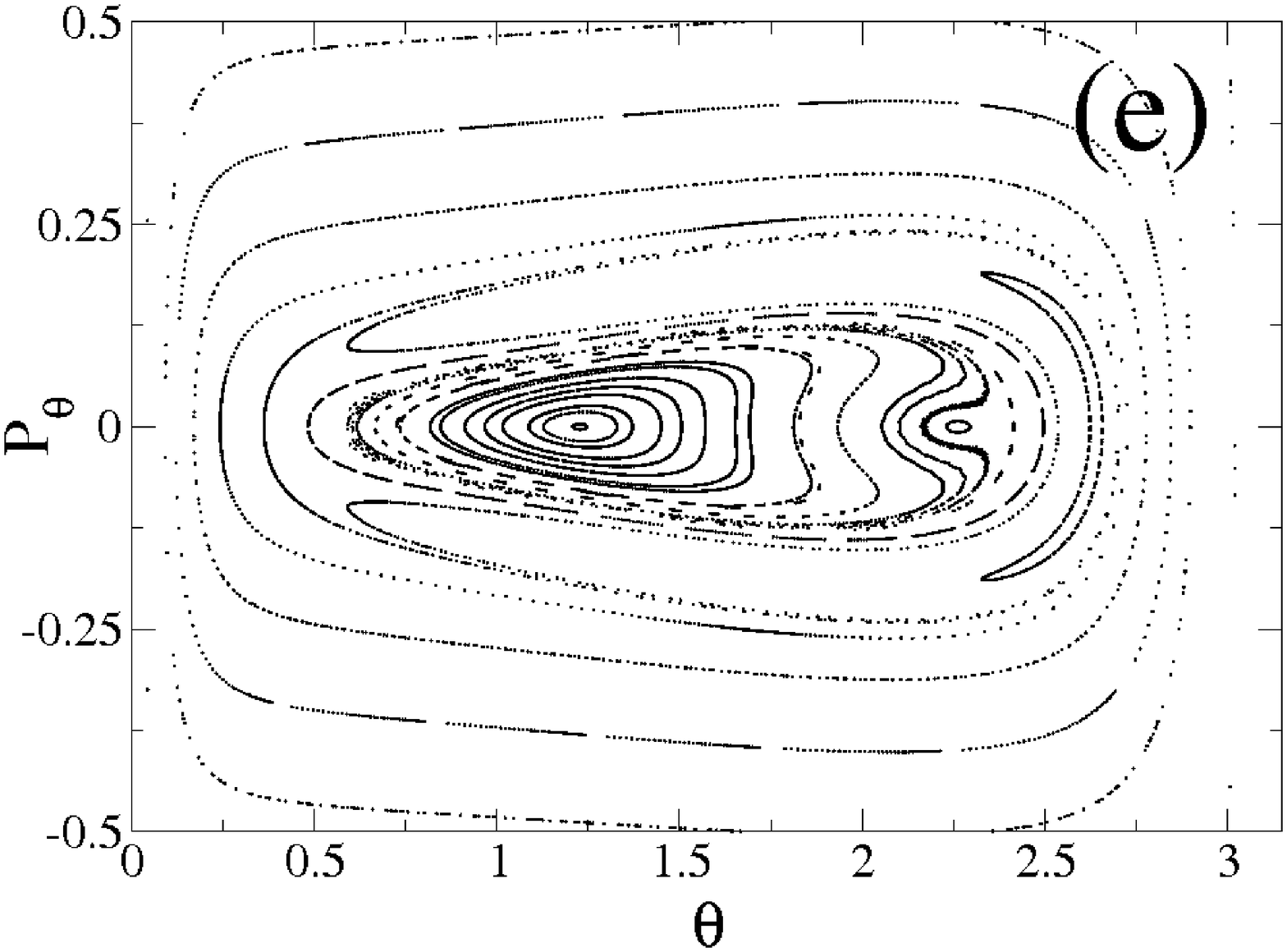}
  \includegraphics[height=45mm,width=45mm,clip]{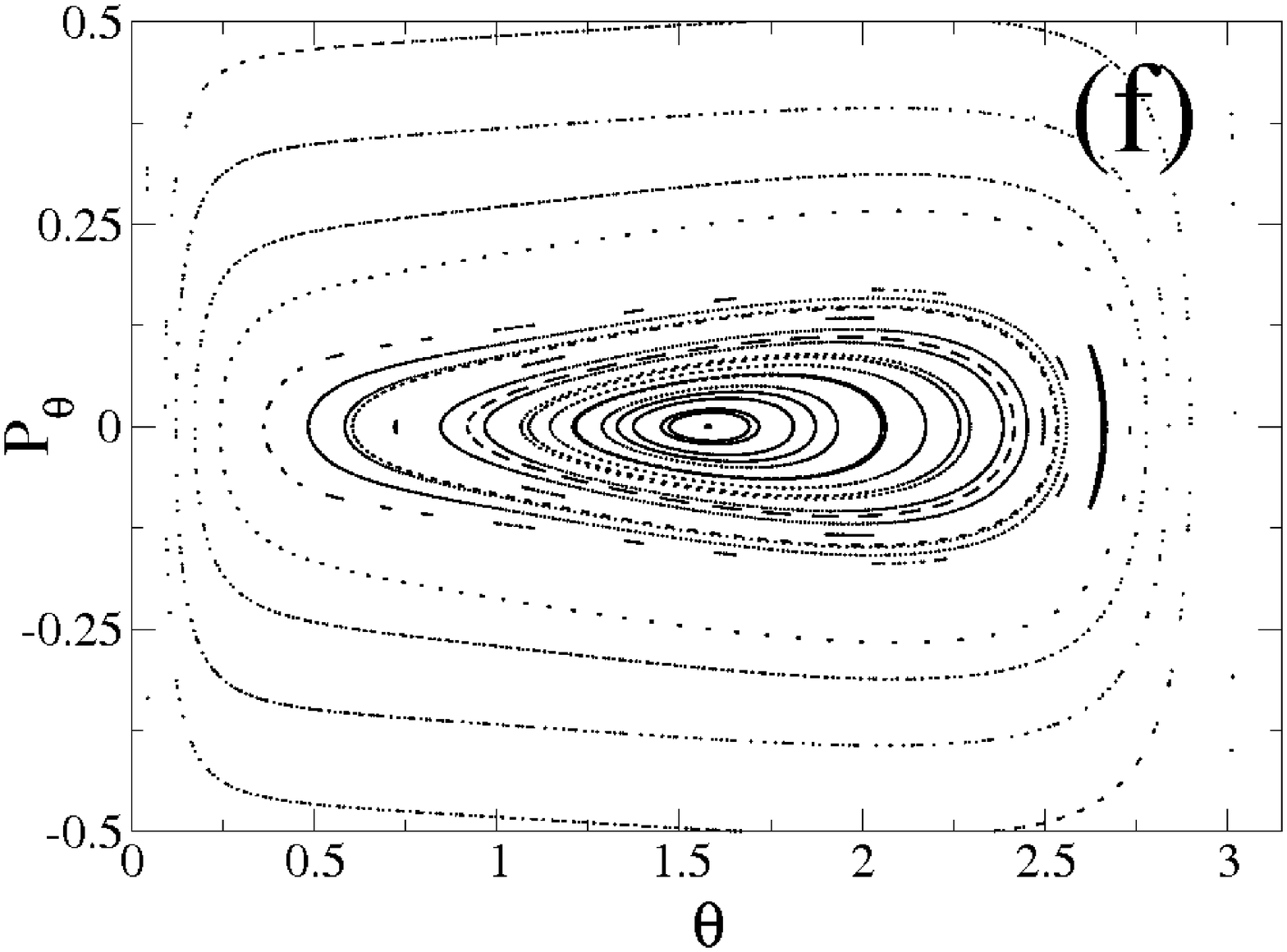}
  \includegraphics[height=45mm,width=45mm,clip]{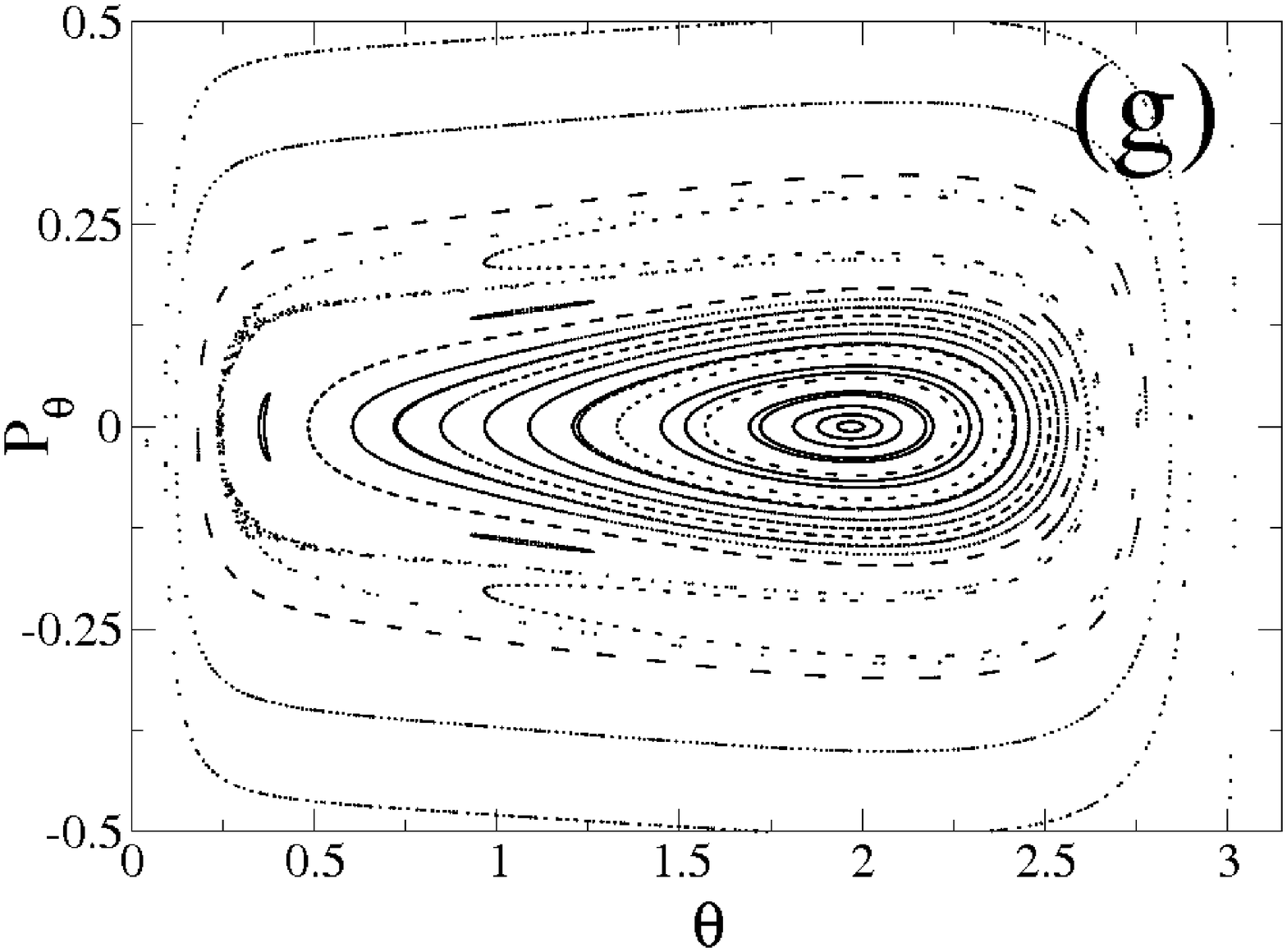}
  \includegraphics[height=45mm,width=45mm,clip]{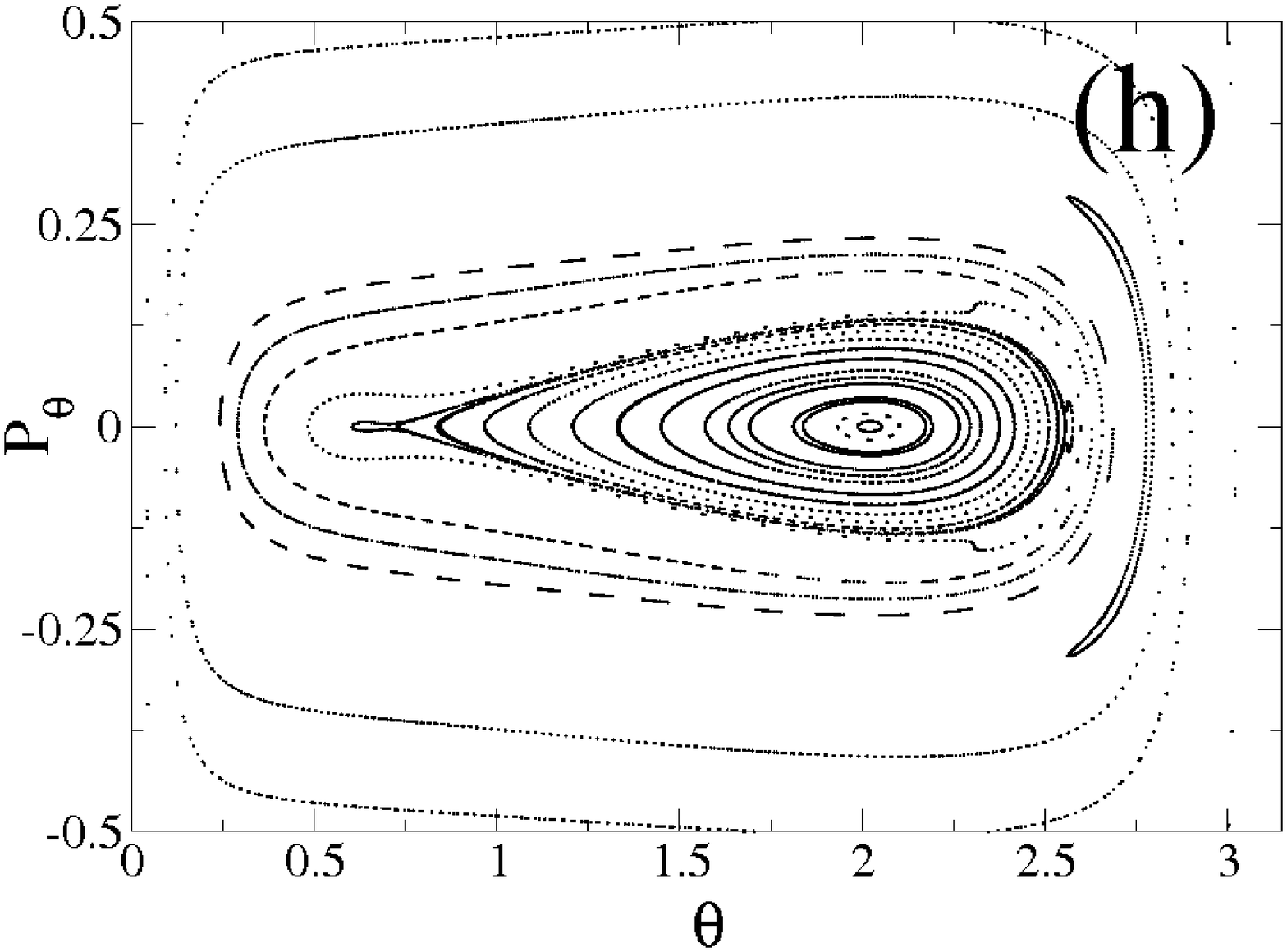}
  \includegraphics[height=45mm,width=45mm,clip]{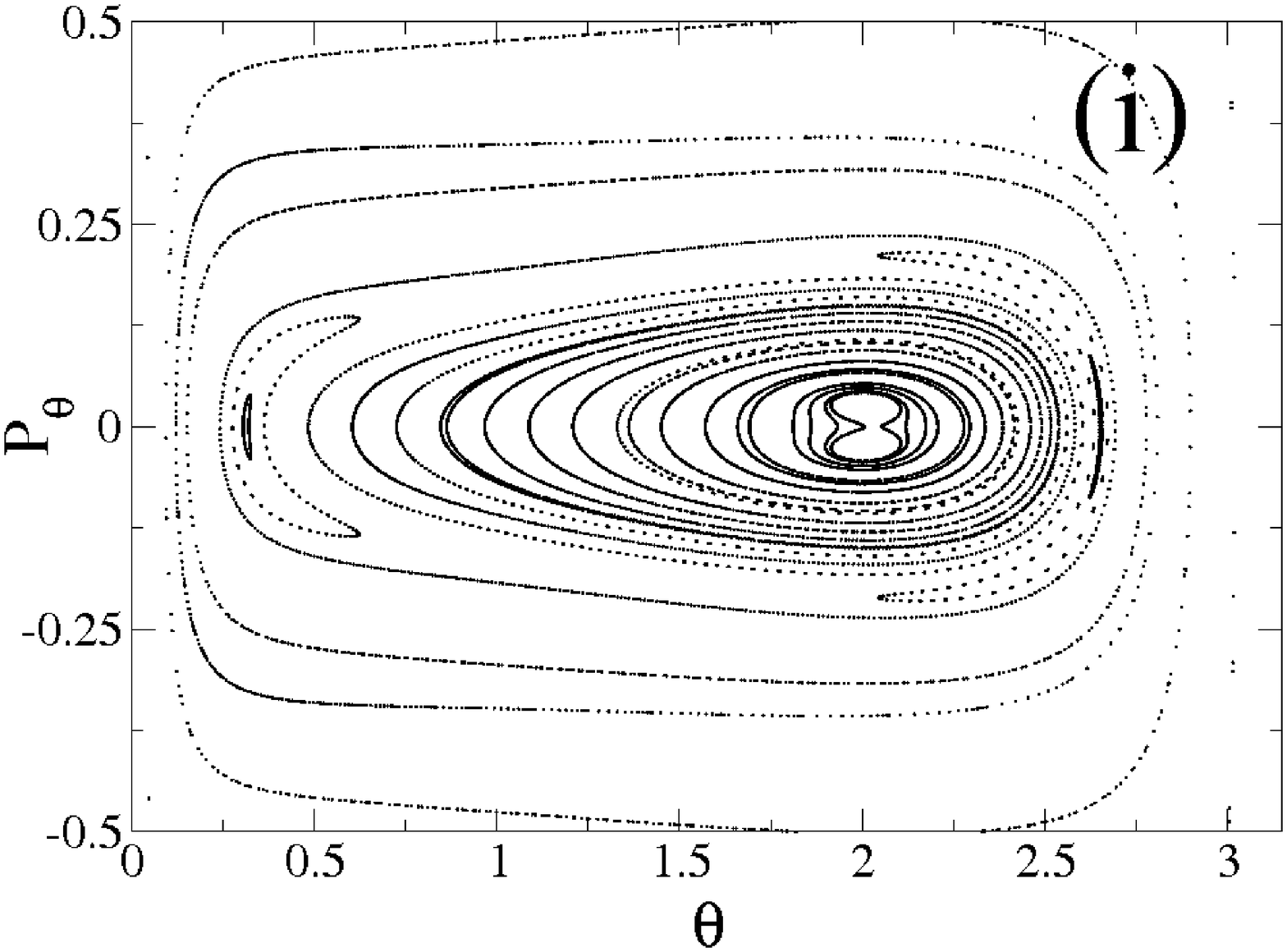}
  \end{center} \caption{Stroboscopic maps for 
  $P_{\psi}=0.086$, $P_{\phi}=0.043$ and $F=20$pN. The
  frequency of the Young's modulus oscillation in each map is: (a)
  $\omega=0.60\omega_{0}$; (b) $\omega=0.82\omega_{0}$; (c)
  $\omega=0.85\omega_{0}$; (d) $\omega=0.90\omega_{0}$; (e)
  $\omega=0.92\omega_{0}$; (f) $\omega=1.00\omega_{0}$; (g)
  $\omega=1.20\omega_{0}$; (h) $\omega=1.60\omega_{0}$; (i)
  $\omega=2.00\omega_{0}$.}
\label{fig1}
\end{figure}

\begin{figure}[ht] 
  \begin{center}
  \includegraphics[height=85mm,width=60mm,clip]{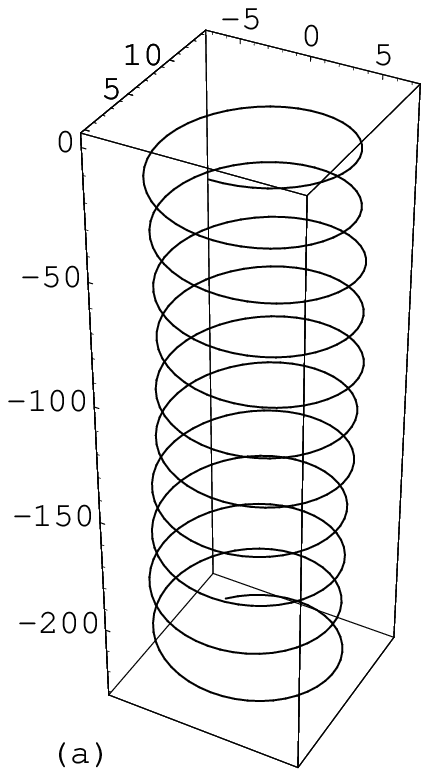}
  \includegraphics[height=85mm,width=60mm,clip]{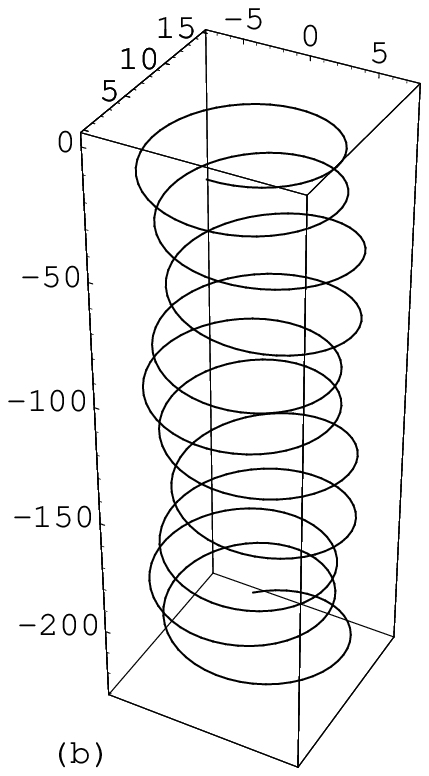}
  \includegraphics[height=85mm,width=60mm,clip]{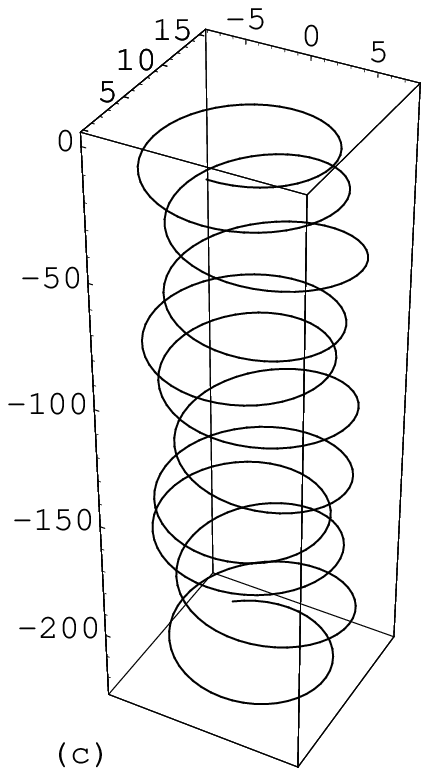}
  \includegraphics[height=85mm,width=60mm,clip]{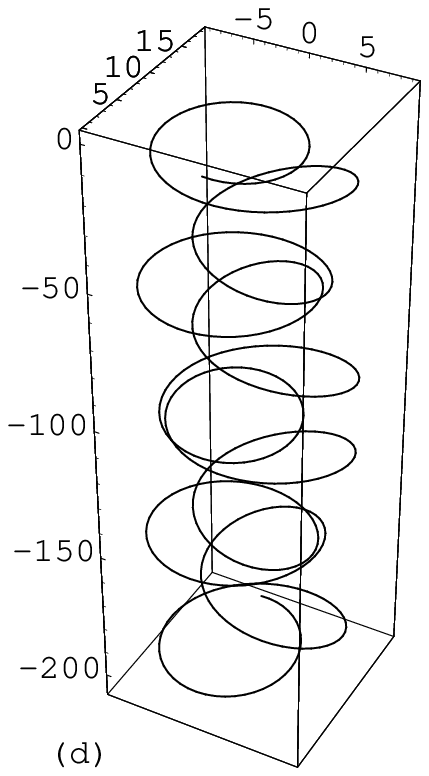}
  \end{center} \caption{Shape evolution of the conformations corresponding 
  to the center of the main island seen in Fig.\ref{fig1}a. 
  (a) $\omega=0.60\omega_{0}$ (Fig.\ref{fig1}a); 
  (b) $\omega=0.82\omega_{0}$ (Fig.\ref{fig1}b);
  (c) $\omega=0.85\omega_{0}$ (Fig.\ref{fig1}c);
  (d) $\omega=0.92\omega_{0}$ (Fig.\ref{fig1}e).}
\label{fig2}
\end{figure}

\begin{figure}[ht] 
  \begin{center}
  \includegraphics[height=85mm,width=60mm,clip]{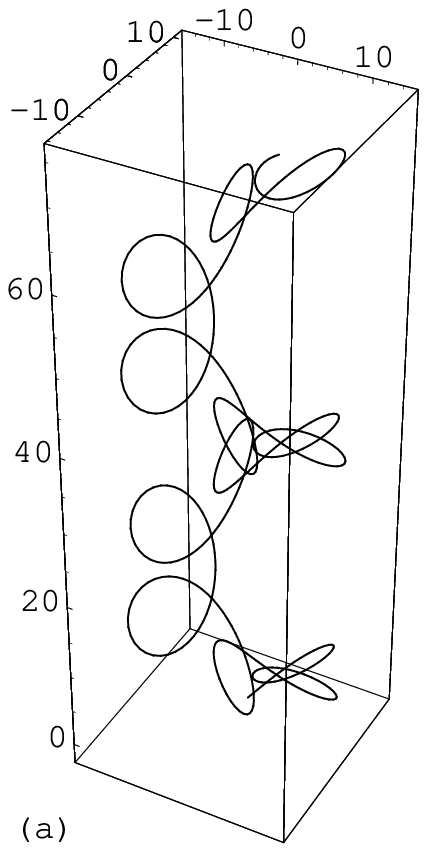}
  \includegraphics[height=85mm,width=60mm,clip]{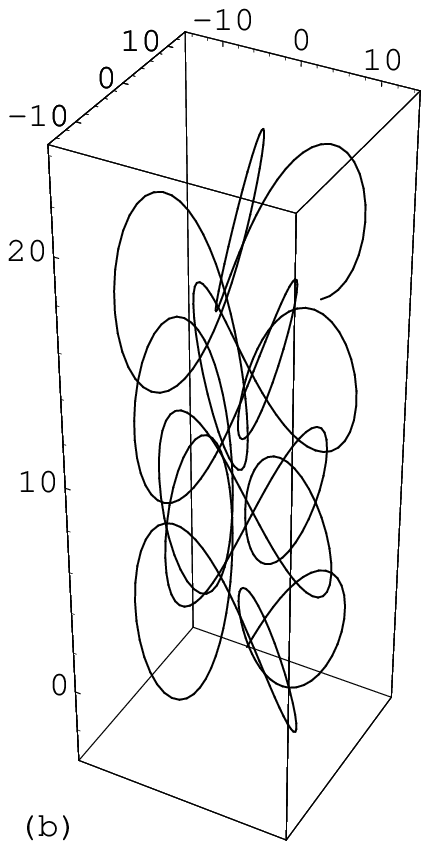}
  \includegraphics[height=85mm,width=60mm,clip]{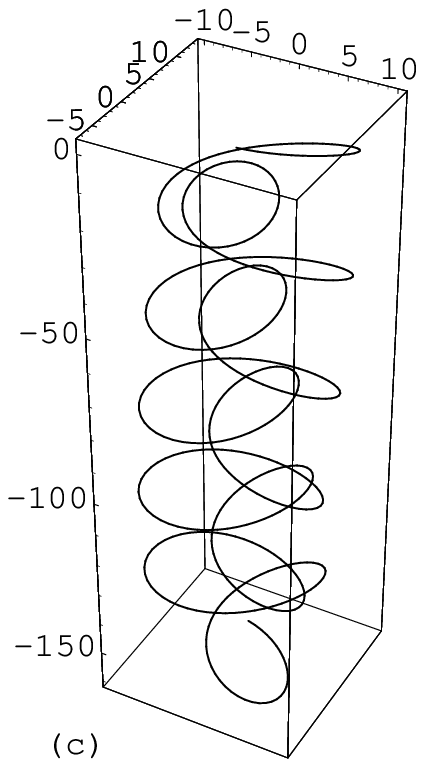}
  \includegraphics[height=85mm,width=60mm,clip]{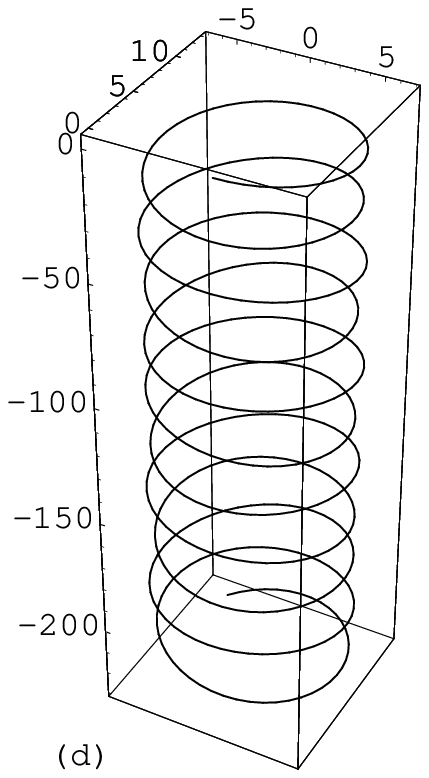}
  \end{center} \caption{Shape evolution of the conformations corresponding 
  to the center of the left island that appears when $\omega > 0.81 \omega_0$
  (Fig.\ref{fig1}b). 
  (a) $\omega=0.82\omega_{0}$ (Fig.\ref{fig1}b);
  (b) $\omega=0.85\omega_{0}$ (Fig.\ref{fig1}c);
  (c) $\omega=\omega_{0}$ (Fig.\ref{fig1}f);
  (d) $\omega=1.60\omega_{0}$ (Fig.\ref{fig1}h).}
\label{fig3}
\end{figure}

\begin{figure}[ht] 
  \begin{center}
  \includegraphics[height=80mm,width=43mm,clip]{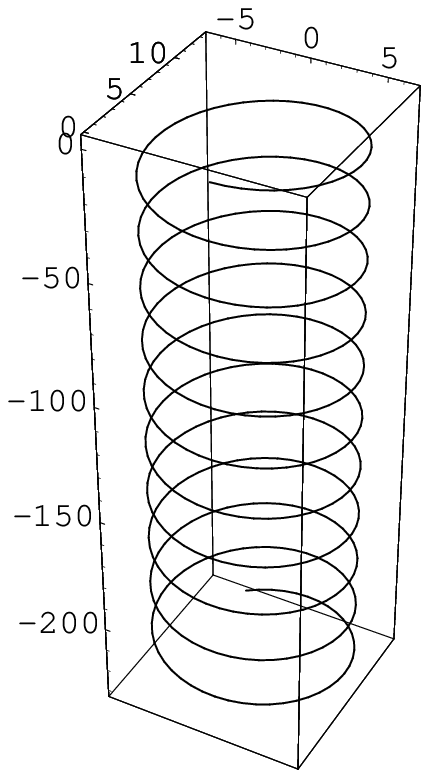}
  \includegraphics[height=80mm,width=43mm,clip]{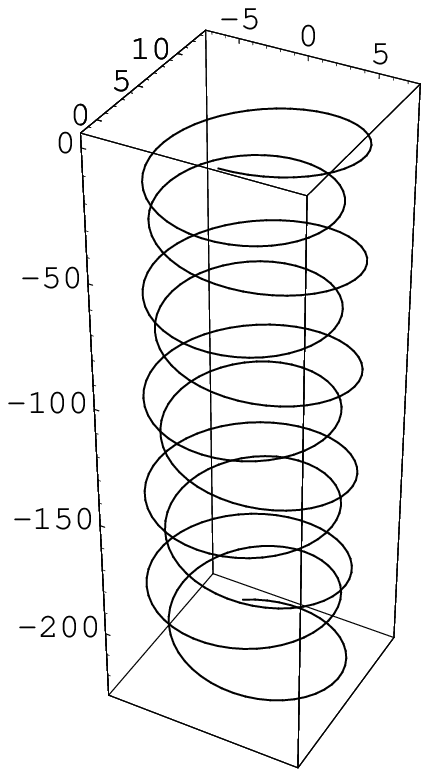}
  \includegraphics[height=80mm,width=43mm,clip]{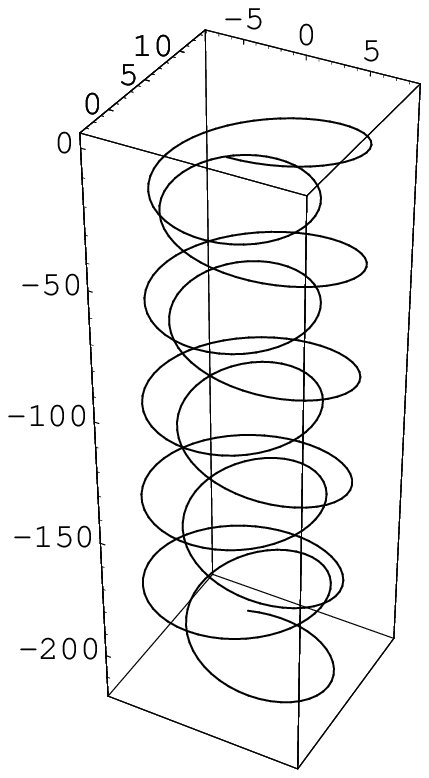} 
  \end{center}
  \caption{Tridimensional shape of the configurations corresponding to
  the center island that appears in the stroboscopic maps for
  $\omega=\omega_0$ and different $\alpha$ (stroboscopic maps not
  shown). From left to right, homogeneous case, $\alpha=0.001$ and
  $\alpha=0.01$.}
\label{fig4}
\end{figure}

\begin{figure}[ht] 
  \begin{center}
  \includegraphics[height=70mm,width=100mm,clip]{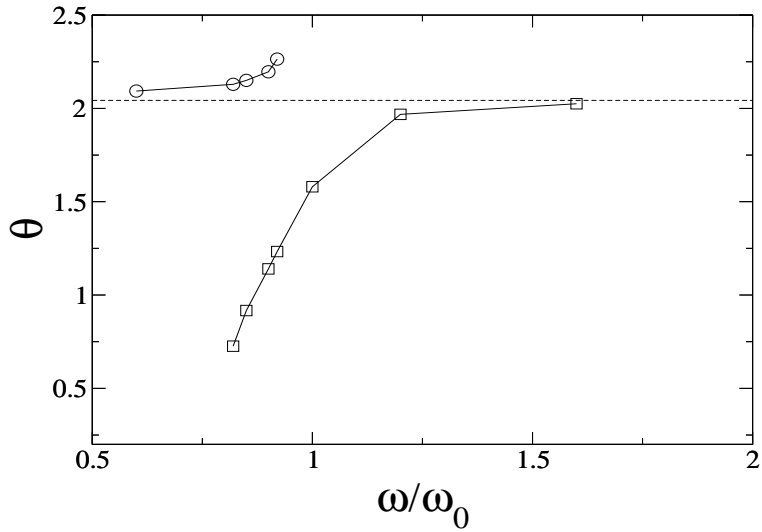}
  \end{center} 
  \caption{Position of the equilibrium points as function of the
  frequency $\omega$, circles corresponding to the center of the main
  island in the stroboscopic maps of Fig~\ref{fig1}a-e, and squares to
  the center of the left island that appears when $\omega > 0.81
  \omega_0$ (Fig\ref{fig1}b-h). The dotted line corresponds to
  $\theta_0$, homogeneous case equilibrium point position.}
\label{fig5}
\end{figure}

Since the coefficients of the terms $(\theta-\theta_0)^n$ have
 the same order of magnitude, it is necessary to check if
$(\theta-\theta_0)<<1$ for all $\theta(s)$, i.e., along the rod. We
found that for the frequency $\omega$ of the Young's
modulus far from the resonance ($|\omega-\omega_0| >> 0$),
the solutions corresponding to the equilibrium points have
$(\theta(s)-\theta_0)_{\rm MAX}\simeq0.03$, where the subscript MAX
means ``maximum value for all $s$''. But at the resonance,
$\omega=\omega_0$, the solution corresponding to the equilibrium point
has $(\theta(s)-\theta_0)_{\rm MAX}\simeq0.5$.
$(\theta(s)-\theta_0)_{\rm MAX}$ becomes even larger than 0.5 if we
consider the solution related to the new equilibrium point (the new
island that appeared in the map displayed in the Fig~\ref{fig1}b).

Therefore, the perturbative method of the dynamical 
systems theory is not applicable to analyzing this case.

\newpage  

\end{document}